\documentclass[sigconf]{acmart}
\makeatletter
\def\@ACM@checkaffil{
    \if@ACM@instpresent\else
    \ClassWarningNoLine{\@classname}{No institution present for an affiliation}%
    \fi
    \if@ACM@citypresent\else
    \ClassWarningNoLine{\@classname}{No city present for an affiliation}%
    \fi
    \if@ACM@countrypresent\else
        \ClassWarningNoLine{\@classname}{No country present for an affiliation}%
    \fi
}
\makeatother
\usepackage{caption}
\usepackage{amsthm}
\usepackage{enumitem}
\usepackage{subcaption}
\usepackage{tikz}
\usepackage{pgfplots} 
\pgfplotsset{compat=1.16} 
\usepackage{balance}
\usepackage{bigstrut,array,multirow,tabularx}
\usepackage{caption}
\usepackage{pifont}
\newcommand{\cmark}{\ding{51}}%
\newcommand{\xmark}{\ding{55}}%
\usepackage{dsfont}
\usepackage{mathtools}

\usepackage{algorithm}
\usepackage{algorithmic}
\usepackage{makecell} 

\definecolor{lightblue}{RGB}{0.93,0.95,1.0} 
\definecolor{lightred}{RGB}{1.0,0.93,0.93} 
\AtBeginDocument{%
  \providecommand\BibTeX{{%
    \normalfont B\kern-0.5em{\scshape i\kern-0.25em b}\kern-0.8em\TeX}}}
\copyrightyear{2024}
\acmYear{2024}
\setcopyright{rightsretained}
\acmConference[SIGIR '24]{Proceedings of the 47th International ACM SIGIR Conference on Research and Development in Information Retrieval}{July 14--18, 2024}{Washington, DC, USA}
\acmBooktitle{Proceedings of the 47th International ACM SIGIR Conference on Research and Development in Information Retrieval (SIGIR '24), July 14--18, 2024, Washington, DC, USA}
\acmDOI{10.1145/3626772.3657916}
\acmISBN{979-8-4007-0431-4/24/07}

\acmConference[SIGIR '24]{the 47th International ACM SIGIR Conference on Research and Development in Information Retrieval}{July 14--18,
  2024}{Washington D.C, USA}
\acmPrice{15.00}
\acmISBN{978-1-4503-XXXX-X/18/06}

\begin{document}

\title{Turbo-CF: Matrix Decomposition-Free Graph Filtering\\ for Fast Recommendation}

\author{Jin-Duk Park}
\affiliation{%
  \institution{Yonsei University}
  \city{Seoul}
  \country{Republic of Korea}}
\email{jindeok6@yonsei.ac.kr}

\author{Yong-Min Shin}
\orcid{1234-5678-9012}
\affiliation{%
  \institution{Yonsei University}
  \city{Seoul}
  \country{Republic of Korea}
}
\email{jordan3414@yonsei.ac.kr}

\author{Won-Yong Shin}
\affiliation{%
  \institution{Yonsei University}
  \city{Seoul}
  \country{Republic of Korea}
}
\email{wy.shin@yonsei.ac.kr}
\authornote{Corresponding author}

\renewcommand{\shortauthors}{Jin-Duk, et al.}

\begin{abstract}
 A series of \textit{graph filtering (GF)}-based collaborative filtering (CF) showcases state-of-the-art performance on the recommendation accuracy by using a low-pass filter (LPF) without a training process. However, conventional GF-based CF approaches mostly perform \textit{matrix decomposition} on the item--item similarity graph to realize the ideal LPF, which results in a non-trivial computational cost and thus makes them less practical in scenarios where rapid recommendations are essential. In this paper, we propose \textsf{Turbo-CF}, a GF-based CF method that is both \textit{training-free} and \textit{matrix decomposition-free}. \textsf{Turbo-CF} employs a \textit{polynomial graph filter} to circumvent the issue of expensive matrix decompositions, enabling us to make full use of modern computer hardware components (\textit{i.e.}, GPU). Specifically, \textsf{Turbo-CF} first constructs an item--item similarity graph whose edge weights are effectively regulated. Then, our own polynomial LPFs are designed to retain only low-frequency signals without explicit matrix decompositions. We demonstrate that \textsf{Turbo-CF} is \textit{extremely fast yet accurate}, achieving a runtime of \textbf{\underline{less than 1 second}} on real-world benchmark datasets while achieving recommendation accuracies comparable to best competitors.

\end{abstract}


\maketitle

\section{Introduction}
\label{section 1}

Recommender systems have significantly impacted various industries, including e-commerce (\textit{e.g.}, Amazon \cite{linden2003amazon}) and content streaming services (\textit{e.g.}, YouTube \cite{covington2016deep} and Netflix \cite{gomez2015netflix}), enabling with personalized recommendations. Central to these systems is collaborative filtering (CF), which predicts a user's preferences based on the historical user--item interactions \cite{wang2019neural,he2020lightgcn,wang2020disentangled,mao2021ultragcn, zheng2018spectral,berg2017graph,he2016fast}.
Notably, CF techniques based on graph convolutional networks (GCNs) have been shown to achieve state-of-the-art recommendation performance by aggregating the information from high-order neighbors through message passing \cite{jin2020multi, he2020lightgcn, chang2021sequential, wu2019session}.

On one hand, it is often the case where the speed at which recommender systems update their models becomes increasingly crucial, mainly due to two key factors. First, user preferences are not static; they tend to evolve rapidly due to various influences such as trends, personal circumstances, and exposure to new content \cite{ju2015using,pereira2018analyzing, chang2021sequential,wu2017modeling}. In this case, recommender systems must be agile enough to reflect such evolving preferences. Second, the influx of new data is often very high, with users constantly interacting with content and services \cite{wu2017modeling,hewanadungodage2017gpu,xu2014taxi}. For example, social media platforms such as Gowalla and Facebook may require quick model updates due to the fast pace of user interactions and content generation \cite{pereira2018analyzing,wu2017modeling}. Therefore, these factors result in the need to build an efficient and prompt system \cite{hewanadungodage2017gpu,ju2015using,goldberg2001eigentaste}. Such a fast adaptation ensures that the recommendations remain relevant and in sync with current user interests and behaviors, thereby enhancing user satisfaction and engagement with the underlying system \cite{goldberg2001eigentaste,toan2018diversifying,ju2015using,pereira2018analyzing, chang2021sequential,wu2017modeling}.

On the other hand, most of existing GCN-based CF methods \cite{berg2017graph, ying2018graph,zheng2018spectral,wang2020disentangled,wang2019neural,he2020lightgcn} have their inherent limitations, notably their reliance on extensive offline training \cite{chi2022long, Jeunen2019Revisiting}. Recently, a series of graph filtering (GF) methods have emerged as another CF technique since they can naturally alleviate this problem with their training-free nature. However, conventional GF-based CF methods suffer from a high computational cost during matrix decomposition to realize the ideal low-pass filter (LPF), hindering their potential for real-time applications~\cite{shen2021powerful,liu2023personalized,choi2023blurring,xia2022fire}.

\begin{figure}[t]
    \centering
    \includegraphics[width=0.7\columnwidth]{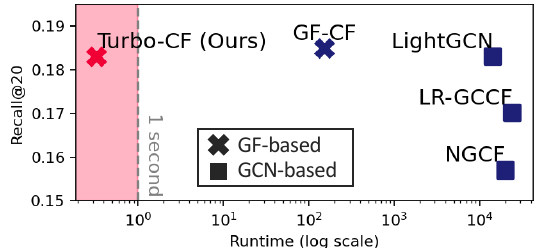}
    \vspace{-3mm}
    \caption{Accuracy versus runtime among \textsf{Turbo-CF} and other benchmark methods on the Gowalla dataset.}
    \label{fig;intro}
\end{figure}

To tackle these practical challenges, we introduce \textsf{Turbo-CF}, a novel GF-based CF method that is both \textit{training-free}
and \textit{matrix decomposition-free}. By harnessing the computational efficiency of \textit{polynomial graph filters}, \textsf{Turbo-CF} is composed solely of simple matrix operations (\textit{i.e.}, matrix multiplications). Thanks to its computational simplicity, it becomes more straightforward for \textsf{Turbo-CF} to make full use of modern computer hardware components such as GPU. As shown in Figure \ref{fig;intro}, we demonstrate that our \textsf{Turbo-CF} is \textit{extremely fast yet sufficiently accurate}, showing a runtime of  \textbf{\underline{less than 1 second}} on real-world benchmark datasets ({\it e.g.}, Gowalla and Yelp) while still showing competitive performance compared to other benchmark CF methods. For reproducibility, the source code is available at https://github.com/jindeok/Turbo-CF.

\section{Preliminary}
\label{section 2}
Given a weighted graph $G = (V, E)$, the Laplacian matrix $L$ of $G$ is $L=D-A$, where $D=\text{diag}(A{\bf 1})$ is the degree matrix for the all-one vector ${\bf 1}\in\mathbb{R}^{|V|}$ and $A$ is the adjacency matrix. A graph signal is represented as $\mathbf{x} \in \mathbb{R}^{|V|}$, where $x_i$ represents the signal strength of node $i$ in ${\bf x}$. The smoothness of $\mathbf{x}$ on $G$ is quantified as $S({\bf x}) = \sum_{i,j}A_{i,j}(x_i-x_j)^2 =  \mathbf{x}^T L \mathbf{x}$, where $A_{i,j}$ is the $(i,j)$-th element of $A$. By the eigen-decomposition $L = U\Lambda U^T$, we can formally define the graph Fourier transform for a graph signal $\mathbf{x}$ as
$\hat{\mathbf{x}} = U^T \mathbf{x}$, where $U\in\mathbb{R}^{|V|\times |V|}$ is the matrix whose columns correspond to a set of eigenvectors of $L$. Now, we are ready to formally define the graph filter and graph convolution as follows:
\begin{definition}{(Graph filter)~\cite{shuman2013emerging, shen2021powerful, xia2022fire,ortega2018graph}}
    Given a graph Laplacian matrix $L$, a graph filter $H(L)\in\mathbb{R}^{|V|\times |V|}$ is defined as 
\begin{equation}
H(L) = U \text{diag}(h(\lambda_1), \ldots, h(\lambda_{|V|})) U^T,
\end{equation}
where $h:\mathbb{C} \rightarrow \mathbb{R}$ is the frequency response function that maps eigenvalues $\{\lambda_1,\cdots,\lambda_{|V|}\}$ of $L$ to $\{h(\lambda_1),\cdots,h(\lambda_{|V|}\}$.
\end{definition}
\begin{definition}{(Graph convolution)~\cite{shuman2013emerging, shen2021powerful, xia2022fire}}
\label{graph_conv_def}
    The convolution of a graph signal $\mathbf{x}$ and a graph filter $H(L)$ is given by
\begin{equation}
    H(L) \mathbf{x} = U \text{diag}(h(\lambda_1), \ldots, h(\lambda_{|V|})) U^T\mathbf{x}.
\end{equation}
\end{definition}

\section{Methodology}
\subsection{Motivation and Challenge}
Suppose that $\mathcal{U}$ and $\mathcal{I}$ are the sets of all users and all items, respectively, in recommender systems dealing with user--item ratings. Then, conventional GF-based CF methods \cite{shen2021powerful,xia2022fire,liu2023personalized} initially construct an item--item similarity graph as
\begin{equation}
\begin{aligned}
    \tilde{P} = \tilde{R}^T\tilde{R}; \tilde{R} = D^{-1/2}_URD^{-1/2}_I.
\end{aligned}
\end{equation}
Here, $R \in \mathbb{R}^{|\mathcal{U}| \times |\mathcal{I}|}$ is the rating matrix; $\tilde{R}$ is the normalized rating matrix; $D_U=\text{diag}(R\mathbf{1})$ and $D_I = \text{diag}(\mathbf{1}^TR)$; and $\tilde{P}$ is the adjacency matrix of the item--item similarity graph.
GF-based CF methods \cite{liu2023personalized,shen2021powerful,xia2022fire} typically employ both linear and ideal LPFs. As the representative work, GF-CF~\cite{shen2021powerful} performs graph convolution as ${\bf s}_u = {\bf r}_u(\tilde{P} + \alpha D^{-1/2}_U\bar{U}\bar{U}^TD^{-1/2}_I)$, where ${\bf s}_u \in \mathbb{R}^{|\mathcal{I}|}$ is the predicted preferences for user $u$; ${\bf r}_u \in \mathbb{R}^{|\mathcal{I}|}$ is the ratings of $u$, serving as graph signals to be smoothed; $\bar{U}\in \mathbb{R}^{|\mathcal{I}| \times k}$ is the top-$k$ singular vectors of $\tilde{R}$; $\tilde{P}$ is the linear LPF; $D^{-1/2}_U\bar{U}\bar{U}^TD^{-1/2}_I$ is the ideal LPF of $\tilde{P}$; and $\alpha$ is a hyperparameter balancing between these two filters. 

Although such GF-based CF methods often employ an ideal LPF, they pose a critical challenge: they necessitate matrix decomposition to acquire $\bar{U}$, incurring substantial computational costs whose complexity is typically $O(|\mathcal{I}|^3)$. This leads to a natural question: \textit{"how can we bypass the problem of matrix decomposition without losing the recommendation accuracy in GF-based CF?"}. To answer this question, we propose \textsf{Turbo-CF}, which effectively solves this challenge using \textit{polynomial graph filters}.

\subsection{Proposed Method: \textsf{Turbo-CF}}
\label{sec 3.2}
\subsubsection{Graph construction}\label{sec 3.2.1}
We describe how to construct an item--item similarity graph with $R$ for GF. Unlike conventional GF-based CF methods that symmetrically normalize $R$ along the user/item axis~\cite{shen2021powerful,xia2022fire,liu2023personalized}, we adopt \textit{asymmetric} normalization on $R$ to regularize the popularity of users/items before calculating $\tilde{P}$, which is formulated as
\begin{equation}
\begin{aligned}
    \tilde{P} = \tilde{R}^T\tilde{R}; \tilde{R} = D^{-\alpha}_URD^{\alpha-1}_I.
\end{aligned}   
\end{equation}
Here, $\alpha \in [0,1]$ is the hyperparameter to control the normalization along users/items. Increasing $\alpha$ weakens the effect of popular users (\textit{i.e.}, high-degree users) while strengthening the effect of popular items. Next, according to the type of graph filter we use based on $\tilde{P}$, the corresponding filtered signals may be over-smoothed or under-smoothed, depending on the intensity of connections between nodes in $\tilde{P}$. Thus, we aim to adjust the edge weights differently depending on the graph filter. To this end, we present an additional adjustment process for the graph $\tilde{P}$ by using the Hadamard power. Finally, the adjusted graph $\bar{P}$ is calculated as
\begin{equation}
    \bar{P}=\tilde{P}^{\circ s},
\end{equation} 
where $s$ is the adjustment parameter that can be tuned based on the validation set. We empirically show that properly adjusting the edge weights in the two graphs ($\tilde{R}$ and $\bar{P}$) substantially improves the recommendation accuracy even without the costly matrix decomposition, which will be verified in Section \ref{sec4.5}.

\subsubsection{Polynomial GF}
\begin{figure}[t]
    \centering
    \includegraphics[width=0.95\columnwidth]{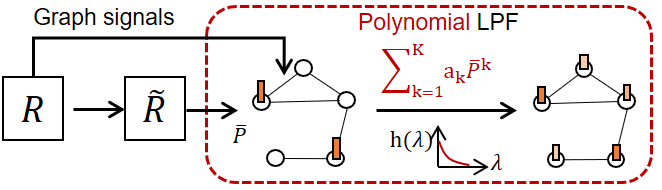}
    \vspace{-3mm}
    \caption{The schematic overview of \textsf{Turbo-CF}. The graph signals are smoothed by using polynomial LPFs in \textsf{Turbo-CF}.}
    \label{overview}
\end{figure}

We now specify how to perform GF based on $\bar{P}$. To bypass the high computation overhead of matrix decompositions in GF, we make use of {\it polynomial} graph filters. We denote the normalized Laplacian matrix of $\bar{P}$ as $\tilde{L} = I - \bar{P}$. Then, since $\tilde{L}$ is a symmetric positive semi-definite matrix, there exists orthogonal eigen-decomposition $\tilde{L} = \tilde{U}\Lambda \tilde{U}^T$, where $\tilde{U}\tilde{U}^T=\tilde{U}^T\tilde{U} =I$. Thanks to $\tilde{U}$ being an orthogonal matrix, we can still manipulate the eigenvalues by the matrix polynomial of $\tilde{L}$ (e.g., $\tilde{L}^2 = \tilde{U}\Lambda^2\tilde{U}^T$) without the need for costly matrix decomposition. The {\it matrix decomposition-free} polynomial graph filter can be expressed as
\begin{equation}
\label{matrix_polynomial}
    \sum_{k=1}^K{a_k}\bar{P}^k,
\end{equation}
where $a_{k}$ is the coefficient of a matrix polynomial and $K$ is the maximum order of the matrix polynomial bases. We establish the following theorem, which provides a closed-form solution to the frequency response function of a polynomial graph filter.
\begin{theorem} 
    \label{thm_poly}
     The matrix polynomial $\sum_{k=1}^K{a_k}\bar{P}^k$ is a graph filter for graph $\bar{P}$ with the frequency response function of $h(\lambda) = \sum_{k=1}^K{a_k}(1 - \lambda)^k$.
\end{theorem}
The proof of Theorem \ref{thm_poly} is omitted due to page limitations. Interestingly, Theorem \ref{thm_poly} implies that we can design arbitrary LPFs by deciding proper coefficients of polynomials, depending on the characteristics of a given task or dataset. Based on the polynomial graph filter in Eq. \eqref{matrix_polynomial}, \textsf{Turbo-CF} can is formalized as follows:
\begin{equation}
    \label{Turbo-CF}
    \textbf{s}_{u} = {\mathbf r}_{u}\sum_{k=1}^K{a_{k}}\bar{P}^k,
\end{equation}
where ${\mathbf r}_{u}$ is the $u$-th row of $R$, which will be used as graph signals of user $u$; and $\sum_{k=1}^K{a_{k}}\bar{P}^k$ is the polynomial graph filter for graph $\bar{P}$. Figure \ref{overview} illustrates how the graph signals are smoothed through the polynomial graph filter in \textsf{Turbo-CF}.

\subsubsection{Filter design} \label{sec3.2.3} 
\begin{figure}[t]
    \centering
    \begin{subfigure}[b]{0.27\linewidth}
        \includegraphics[width=\linewidth]{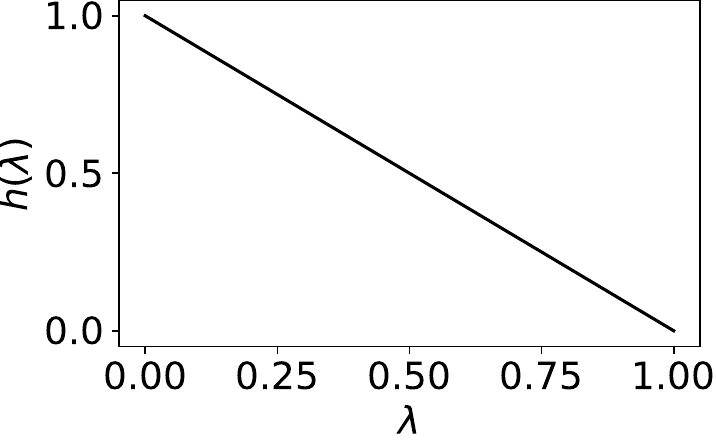}
        \vspace{-7mm}
        \caption{Linear}
        \label{fig:L}
    \end{subfigure}
    \hfill
    \begin{subfigure}[b]{0.27\linewidth}
        \includegraphics[width=\linewidth]{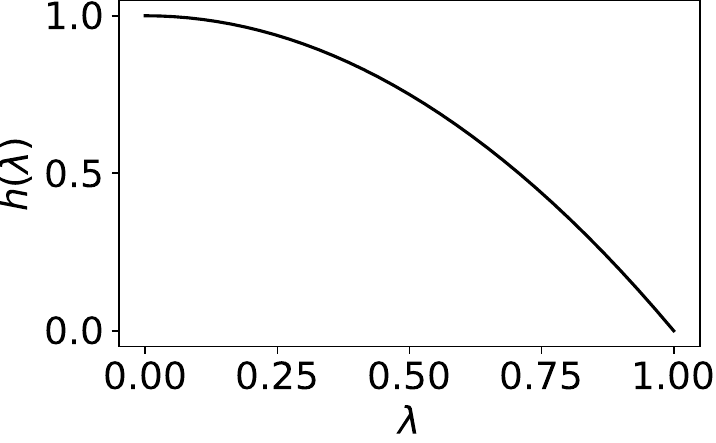}
        \vspace{-7mm}
        \caption{Second-order}
        \label{fig:I}
    \end{subfigure}
    \hfill
    \begin{subfigure}[b]{0.27\linewidth}
        \includegraphics[width=\linewidth]{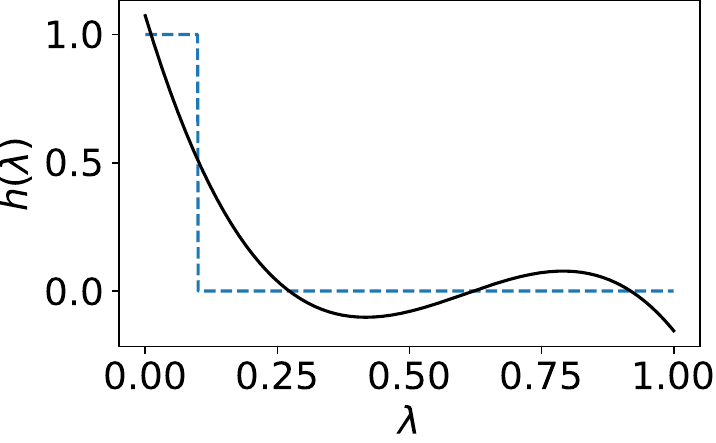}
        \vspace{-7mm}
        \caption{Poly. approx.}
        \label{fig:O}
    \end{subfigure}
\vspace{-2mm}
    \caption{Frequency response functions of three polynomial LPFs. In Figure 3(c), the dotted blue line corresponds to the ideal LPF $h(\lambda) = {\bf 1}_{\lambda\le0.1}$.}
    \label{lpf_plot}
\end{figure}

We note that discovering the optimal polynomial low-pass graph filter in Eq. \eqref{Turbo-CF} involves an extensive search for coefficients $\{a_{k}\}_{k=1}^K$ using a validation set. This process is not ideal as it requires costly recalibration with varying datasets. Additionally, since GF-based CF methods necessitate loading the high-dimensional matrix $\bar{P}$ into memory, handling high-order polynomials (\textit{e.g.}, $K>3$) can cause additional space complexity issues. To this end, as an alternative, we design three polynomial LPFs up to the third order ({\it i.e.}, $K=3$) for GF. Figure \ref{lpf_plot} displays the frequency response functions of the three polynomial LPFs. The explicit forms of the three LPFs derived from Theorem 3.1 are expressed as follows.
\begin{itemize}
\vspace{-2mm}
    \item \textbf{(Linear LPF)} This filter employs the first-order matrix polynomial $\bar{P}$ as a low-pass graph filter. Its frequency response function is $h(\lambda) = 1-\lambda$.
    \item \textbf{(Second-order LPF)} This filter employs the second-order matrix polynomial $2\bar{P}-\bar{P}^2$ as a low-pass graph filter. Its frequency response function is $h(\lambda) = 1-\lambda^2$.

    \item \textbf{(Polynomial approximation to ideal LPF)} Previous studies \cite{shen2021powerful,xia2022fire,liu2023personalized} have proven that utilizing an ideal LPF (\textit{i.e.}, $h(\lambda) = {\bf 1}_{\lambda\le\tau}$ with cutoff frequency $\tau$) alongside a linear LPF is indeed beneficial in improving the recommendation accuracy. In this context, we aim to gain a similar effect to the case of using such an ideal LPF while avoiding the computational demands of matrix decomposition. To this end, we numerically approximate the ideal LPF using a polynomial function. Precisely, a linear LPF combined with the ideal LPF can be formulated as
    $\bar{P} + \beta \hat{f}_{\tau}(\bar{P})$, where $\beta$ is a hyperparameter and $\hat{f}_{\tau}(\bar{P})$ is the approximated ideal LPF with cutoff frequency $\tau$. In our study, we employ the third-order polynomial function to balance the approximation quality and computational burden. We numerically solve a non-linear least squares problem to find the coefficients $\{a_{k}\}_{k=1}^3$ in Eq. \eqref{matrix_polynomial} for the approximation of $h(\lambda) = {\bf 1}_{\lambda\le\tau}$. For instance, as shown in Figure \ref{lpf_plot}(c), we can pre-compute the coefficients of polynomials as $\hat{f}_{\tau}(\bar{P}) = -\bar{P}^3+10\bar{P}^2-29\bar{P}$ for the ideal LPF with $\tau=0.1$.  
\end{itemize}
\vspace{-1mm}
 We note that, besides the aforementioned three LPFs, one can design any LPFs by setting $\{a_k\}_{k=1}^K$ based on one's own design choice. It is also worth noting that, as shown in Eq. \eqref{Turbo-CF}, the polynomial low-pass graph filters can be implemented through simple matrix calculations, which allows us to more effectively leverage well-optimized machine learning and computation frameworks such as PyTorch \cite{paszke2019pytorch} and CUDA \cite{sanders2010cuda} via parallel computation. Thanks to such rapid computation of polynomial low-pass graph filters via our \textsf{Turbo-CF} method, the optimal filter for a given dataset can also be readily found using the validation set.
\vspace{-1mm}

\section{Experimental results and Analyses}
\begin{table}[t]
\small
  \captionsetup{skip=2pt}
  \caption{The statistics of three datasets.}
  \begin{tabular}{ccccccl}
    \toprule
    Dataset & \# of users & \# of items & \# of interactions & Density \\ 
    \midrule
    Gowalla & 29,858 & 40,981 & 1,027,370 & 0.084\%\\
    Yelp& 31,668 & 38,048 & 1,561,406 &  0.130\%\\
    Amazon-book & 52,643 & 91,599& 2,984,108 & 0.062\% \\
  \bottomrule
\end{tabular}
\vspace{0mm}
\label{table:datasets}
\end{table}


\subsection{Experimental Settings}
\noindent\textbf{Datasets.} We carry out experiments on three datasets: Gowalla, Yelp, and Amazon-book. Statistics of the three datasets are summarized in Table \ref{table:datasets}. 

\noindent\textbf{Benchmark methods.} We compare \textsf{Turbo-CF} with eight benchmark CF methods, including matrix factorization-based (MF-BPR \cite{rendle2009bpr} and NeuMF \cite{he2017neural}), generative model-based (Multi-VAE \cite{liang2018variational} and DiffRec \cite{wang2023diffusion}), GCN-based (NGCF \cite{wang2019neural}, LightGCN \cite{he2020lightgcn}, LR-GCCF \cite{chen2020revisiting}, and GF-based (GF-CF \cite{shen2021powerful}) methods.

\noindent\textbf{Evaluation protocols.} We randomly select 70/20/10\% of the interactions for each user as the training/test/validation sets, where the validation set is used for hyperparameter tuning as well as LPF selection in Section \ref{sec3.2.3}. We use Recall@$K$ and normalized discounted cumulative gain (NDCG@$K$) as our performance metrics, where $K$ is set to 20 by default. 

\noindent\textbf{Implementation details.} To ensure the optimal performance of the benchmark methods used for comparison, we directly quote the results reported \cite{he2020lightgcn,choi2023blurring,xu2018powerful}, except DiffRec \cite{wang2023diffusion} where the results were not reported on the same datasets. All experiments are carried out on a machine with Intel (R) 12-Core (TM) i7-9700K CPUs @ 3.60 GHz and an NVIDIA GeForce RTX A6000 GPU.
\vspace{-1mm}
\subsection{Runtime Comparison}
Table \ref{runtime_table} summarizes the runtime of \textsf{Turbo-CF} using three different graph filters against some of the benchmark methods on the Gowalla dataset as other datasets exhibit similar tendencies. Here, runtime indicates the training time for GCN-based approaches (NGCF and LightGCN) while the processing time for GF-based approaches (GF-CF and \textsf{Turbo-CF}). Our findings are as follows:
\begin{enumerate}[label=(\roman*)]
    \item  In comparison with LightGCN using a lightweight architecture design, \textsf{Turbo-CF} significantly boosts the computational efficiency, achieving up to $\times48,230$ faster runtime without costly model training. 
    
    \item Even compared to GF-CF, \textsf{Turbo-CF} offers $\times510$ faster runtime when the linear LPF is used. This is attributed to the design of \textsf{Turbo-CF} where matrix decomposition is unnecessary and rather much simpler polynomial GF, composed of simple matrix multiplications, is utilized.
\end{enumerate}
\vspace{-2mm}
\begin{table}[t]
\small
\centering
\caption{Runtime of baselines and \textsf{Turbo-CF} using the three different graph filters in Section \ref{sec3.2.3} on the Gowalla dataset.}
\label{runtime_table}
\vspace{-1mm}
\begin{tabular}{lccccc}
\hline
\textbf{} & \textbf{NGCF}   & \textbf{LightGCN}  & \textbf{GF-CF} & \textbf{\textsf{Turbo-CF}} \\ \hline
Runtime & 6h34m15s & 4h1m9s & 2m33s  & \textbf{0.3s}/5.6s/5.8s\\
Training & \textcolor{green}{\cmark}  & \textcolor{green}{\cmark} & \textcolor{red}{\xmark} & \textcolor{red}{\xmark}  \\
\hline
\vspace{-5mm}
\end{tabular}
\end{table}
\subsection{Recommendation Accuracy}
\begin{table}[!t]\centering
\setlength\tabcolsep{5.0pt}
\footnotesize
  \captionsetup{skip=2pt}
  \caption{Performance comparison among \textsf{Turbo-CF} and competitors. The best and second-best performers are highlighted in bold and underline, respectively.}
  \label{main results}
  \begin{tabular}{c|cc|cc|cc}
    \toprule[1pt]
    \multicolumn{1}{c|}{}&\multicolumn{2}{c|}{Gowalla}&\multicolumn{2}{c|}{Yelp}&\multicolumn{2}{c}{Amazon-book}\\
    \cmidrule{1-7}
           Method &   Recall& NDCG&  Recall& NDCG&  Recall& NDCG\\
    \midrule[1pt]
    MF-BPR  &  0.1291 & 0.1109 &   0.0433 & 0.0354 &  0.0250  &0.0190\\
    NeuMF   &  0.1399 & 0.1212 &   0.0451 & 0.0363 &  0.0258  &0.0200\\
    Multi-VAE & 0.1641 & 0.1335 &   0.0584 & 0.0450 &   0.0407 & 0.0315\\
    DiffRec   &  0.1653&  0.1417 &0.0656 & 0.0552  &  OOM  & OOM\\
    NGCF    &  0.1570 & 0.1327 &   0.0579 & 0.0477 &   0.0344 & 0.0263\\
    LightGCN&  0.1830 & \textbf{0.1554} &   0.0649 & 0.0530 &   0.0411 & 0.0315\\
    LR-GCCF&  0.1701 & 0.1452 &   0.0604 & 0.0498 &  0.0375  & 0.0296\\
    GF-CF   &  \textbf{0.1849} & 0.1518  &\textbf{0.0697} & 
    \underline{0.0571}  &  \underline{0.0710}  & \underline{0.0584}\\    
     \cmidrule{1-7}
    \textsf{Turbo-CF}& \underline{0.1835}& \underline{0.1531} & \underline{0.0693} & \textbf{0.0574} & \textbf{0.0730} & \textbf{0.0611}\\
    \bottomrule[1pt]
  \end{tabular}
  \vspace{-1mm}
\end{table}
Table \ref{main results} compares the performance of \textsf{Turbo-CF} and eight competitors. Our findings are as follows:
\begin{enumerate}[label=(\roman*)]
    \item  \textsf{Turbo-CF} consistently achieves competitive performance compared to the state-of-the-art CF methods, while achieving the best performance in NDCG@20 on Yelp and in both metrics on Amazon-book. Specifically, \textsf{Turbo-CF} shows gains up to 4.6\% in terms of NDCG, compared to the second-best performer on the Amazon-book dataset.

    \item The competitive performance of \textsf{Turbo-CF} demonstrates that the polynomial GF is still effective in guaranteeing satisfactory recommendation accuracies, even without using a matrix decomposition-aided ideal LPF in GF-CF.

    \item In short, \textsf{Turbo-CF} is extremely fast yet reliable, making it a strong benchmark for CF-based recommender systems.

\end{enumerate}
\subsection{Effect of Polynomial Graph Filters}

To discover a relationship between accuracy and runtime, Table \ref{filter_table} summarizes the runtime and recall according to three different polynomial graph filters in \textsf{Turbo-CF}. We present three variants of \textsf{Turbo-CF}, namely \textsf{Turbo-CF-1}, \textsf{Turbo-CF-2}, and \textsf{Turbo-CF-3}, which represent \textsf{Turbo-CF} using the linear LPF, second-order LPF, and polynomially approximated LPF to the ideal one, respectively.
 \begin{enumerate}[label=(\roman*)]

    \item The optimal polynomial graph filter in terms of Recall@$20$ is observed differently depending on each dataset. \textsf{Turbo-CF-3}, \textsf{Turbo-CF-2}, and \textsf{Turbo-CF-1} achieve their best performance on Gowalla, Yelp, and Amazon-book, respectively. This indicates that higher-order polynomial LPFs do not always guarantee better performance and using a linear LPF is often sufficient to achieve satisfactory performance.
    
    \item As stated in Section \ref{sec3.2.3}, GF-based CF methods require loading the high-dimensional matrix $\bar{P}$ into memory. Thus, the inclusion of high-order polynomials (\textsf{Turbo-CF-2} and \textsf{Turbo-CF-3}) causes out-of-memory (OOM) issues on larger datasets such as Amazon-book. To relieve this, we can utilize efficient algorithms for matrix multiplication such as the Strassen algorithm \cite{thottethodi1998tuning, huss1996implementation} or submatrix partitioning \cite{goto2008anatomy,choi2020matrix}; this enables us to greatly reduce the memory complexity when large-scale matrix multiplication is involved.
\end{enumerate}
\vspace{-3mm}

\begin{table}[t!]
    \footnotesize
    \centering
    \caption{Runtime and recall according to three different polynomial graph filters in \textsf{Turbo-CF}.}
    \vskip -0.1in
    \begin{tabular}{lcccccc}
    \toprule
    & \multicolumn{2}{c}{\textbf{Gowalla}} & \multicolumn{2}{c}{\textbf{Yelp}} & \multicolumn{2}{c}{\textbf{Amazon-book}}  \\
    \cmidrule(r){2-3} \cmidrule(r){4-5} \cmidrule(r){6-7}
    & Runtime & Recall & Runtime & Recall & Runtime & Recall \\
    \midrule
    \textsf{Turbo-CF-1} & 0.32s & 0.1823 & 0.31s & 0.0689 & 27.9s & 0.0730  \\
    \textsf{Turbo-CF-2} & 5.59s & 0.1740 & 4.80s & 0.0693 & OOM &  OOM \\
    \textsf{Turbo-CF-3} & 5.79s & 0.1835 & 4.82s & 0.0689 & OOM &  OOM \\
    \bottomrule
    \vspace{-5mm}
    \end{tabular}
    \label{filter_table}
\end{table}
\definecolor{linecol1}{rgb}{1.0, 0.5, 0.0}
\definecolor{linecol2}{rgb}{0.1, 0.6, 0.0}
\definecolor{linecol3}{rgb}{0.2, 0.4, 0.8}
\definecolor{linecol4}{rgb}{0.2, 0.1, 0.7}
\begin{figure}[t!]
\vspace{-5mm}
\pgfplotsset{footnotesize,samples=10}
\centering
\begin{tikzpicture}
\begin{axis}[
xmax=0.8,xmin=0.3,ymin= 0.1,ymax=0.2,
xlabel=(a) Effect of $\alpha$., width = 4.0cm, height = 2.7cm, 
xtick={0.3, 0.4, 0.5, 0.6, 0.7, 0.8},ytick={0.1, 0.15, 0.2, 0.3}]
    \addplot+[color=linecol2] coordinates{(0.3,0.1260) (0.4,0.1607) (0.5,0.1741) (0.6,0.1823) (0.7,0.1835) (0.8,0.1813) };
    \addplot+[color=linecol2] coordinates{(0.3,0.1260) (0.4,0.1607) (0.5,0.1741) (0.6,0.1823) (0.7,0.1835) (0.8,0.1813) };
\end{axis}
\end{tikzpicture}
\begin{tikzpicture}
\begin{axis}[
xmax=1,xmin=0.5,ymin= 0.1,ymax=0.2,
xlabel=(b) Effect of $s$., width = 4.0cm, height = 2.7cm,
xtick={0.5, 0.6, 0.7, 0.8, 0.9, 1},ytick={0.1, 0.15, 0.2, 0.3}]
    \addplot+[color=linecol2] coordinates{(0.5,0.1317) (0.6,0.1835) (0.7, 0.1814) (0.8, 0.1800) (0.9, 0.1785) (1, 0.1765) };
    \addplot+[color=linecol2] coordinates{(0.5,0.1317) (0.6,0.1835) (0.7, 0.1814) (0.8, 0.1800) (0.9, 0.1785) (1, 0.1765) };
\end{axis}
\end{tikzpicture}
\vspace{-3mm}
\caption{The effect of two hyperparameters on the Recall@20 for the Gowalla dataset.}
\label{hyper_plot}

\end{figure}
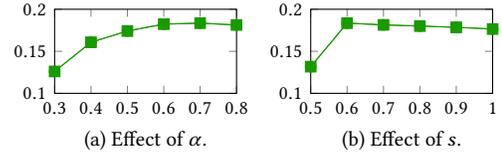

\subsection{Sensitivity Analysis} \label{sec4.5} 
We analyze the impact of the key parameters in \textsf{Turbo-CF}, including two balancing parameters $\alpha$ and $s$ for graph construction, on the recommendation accuracy for the Gowalla dataset as other datasets exhibit similar tendencies. Other parameters are set to the pivot values. First, Figure \ref{hyper_plot}(a) shows that the optimal $\alpha$ is found at $\alpha = 0.7$, which confirms that symmetric normalization $\alpha = 0.5$ is not an optimal choice for ensuring accurate recommendations. Next, Figure \ref{hyper_plot}(b) shows that the optimal $s$ is found at $s=0.6$, not at $s=1$. The sensitivity analysis on $\alpha$ and $s$ validates that properly adjusting the two balancing parameters $\alpha$ and $s$ leads to sufficient gains.

\section{Conclusions}
In this paper, we proposed \textsf{Turbo-CF}, a GF-based CF method that is both \textit{training-free} and \textit{matrix decomposition-free}. \textsf{Turbo-CF} was built upon a computing hardware-friendly \textit{polynomial graph filter} to circumvent the issue of expensive matrix decomposition. Extensive experiments demonstrated (a) the extraordinarily computational efficiency of \textsf{Turbo-CF} with a runtime of less than 1 second on Gowalla and Yelp datasets and (b) the competitive recommendation accuracy of \textsf{Turbo-CF} compared to other benchmark methods.
\label{section 5}

\section*{Acknowledgments}
This research was supported by the National Research Foundation of Korea (NRF) grant funded by the Korea government (MSIT) (No. 2021R1A2C3004345, No. RS-2023- 00220762).








%


\bibliographystyle{ACM-Reference-Format}
\bibliography{sample-base, citation_list}










\end{document}